\documentclass[twocolumn, superscriptaddress]{revtex4-2}
\usepackage{graphicx} % Required for inserting images
\usepackage{amsmath}
\usepackage[braket, qm]{qcircuit}
\usepackage{braket}

\usepackage[percent]{overpic}

\usepackage[export]{adjustbox} % For positioning labels over images

\begin{document}
\title{Fast and Noise-aware Machine Learning Variational Quantum Eigensolver Optimiser}

\author{Akib Karim}
    \email{akib.karim@csiro.au}\affiliation{Data 61, CSIRO, Research Way Clayton 3168, Victoria, Australia}
\author{Shaobo Zhang}\affiliation{School of Physics, University of Melbourne, Parkville 3010, Victoria, Australia}
\author{Muhammad Usman}\affiliation{Data 61, CSIRO, Research Way Clayton 3168, Victoria, Australia}\affiliation{School of Physics, University of Melbourne, Parkville 3010, Victoria, Australia}

\begin{abstract}

The Variational Quantum Eigensolver (VQE) is a hybrid quantum-classical algorithm for preparing ground states in the current era of noisy devices. The classical component of the algorithm requires a large number of measurements on intermediate parameter values that are typically discarded. However, intermediate steps across many calculations can contain valuable information about the relationship between the quantum circuit parameters, resultant measurements, and noise specific to the device. In this work, we use supervised machine learning on the intermediate parameter and measurement data to predict optimal final parameters. Our technique optimises parameters leading to chemically accurate ground state energies much faster than conventional techniques. It requires significantly fewer iterations and simultaneously shows resilience to coherent errors if trained on noisy devices. We demonstrate this technique on IBM quantum devices by predicting ground state energies of H$_2$ for one and two qubits; H$_3$ for three qubits; and HeH$^+$ for four qubits where it finds optimal angles using only modeled data for training.
\end{abstract}

\maketitle
\section{Introduction}

Quantum computing is a rapidly progressing field that offers the potential for advantages in solving specific computationally hard problems at scale such as factoring to break encryption, adversarially robust machine learning, or simulation of quantum systems~\cite{Montanaro2016,Shor1994,West2023}. In particular, the Variational Quantum Eigensolver (VQE) has gained the status of a key quantum algorithm for use in quantum chemistry, and materials science~\cite{Peruzzo2014,Tilly2022} and its variations are able to solve problems in combinatorial optimisation~\cite{farhi2014,yung2024}, quantum compilation~\cite{Khatri2019}, and differential equations~\cite{Elfving2021}.

For quantum chemistry, there exist alternative algorithms such as quantum phase estimation~\cite{kitaev1995} or adiabatic state preparation~\cite{Aspuru-Guzik2005}, but these require deep circuits and are impractical until fault tolerant devices are available~\cite{Cerezo2021}. For noisy devices with a limited number of qubits, VQE is of particular interest due to its resilience to coherent noise~\cite{Peruzzo2014}. VQE is a hybrid classical-quantum algorithm where a classical minimiser finds quantum circuit parameters that minimise the loss function. The circuit parameters are not important as long as the correct state is prepared, therefore, noise that can be corrected by rotating the parameters will naturally be compensated during minimisation, for example calibration error or noise channels that rotate the state coherently~\cite{Fontana2021,Peruzzo2014,Cerezo2021}. In fact, quantum compilation has also been shown to be resilient to incoherent noise since the value of the loss function is not important, only preparing the correct state~\cite{Sharma2020}. The inherent noise resilience makes VQE invaluable for use on current noisy devices. 

Despite highly promising use cases for VQE, a key challenge is that the classical optimisation of variational circuit parameters takes an unpredictable and often large number of iterations whereas only the final iteration is valuable as the solution. This is especially problematic as the optimisation routine may be trapped in a barren plateau where the gradient vanishes exponentially with problem size~\cite{Arrasmith2021,Tilly2022}. 

\begin{figure*}
    \centering
    \begin{overpic}[width=0.85\textwidth,trim={0 1cm 0 1cm},clip]{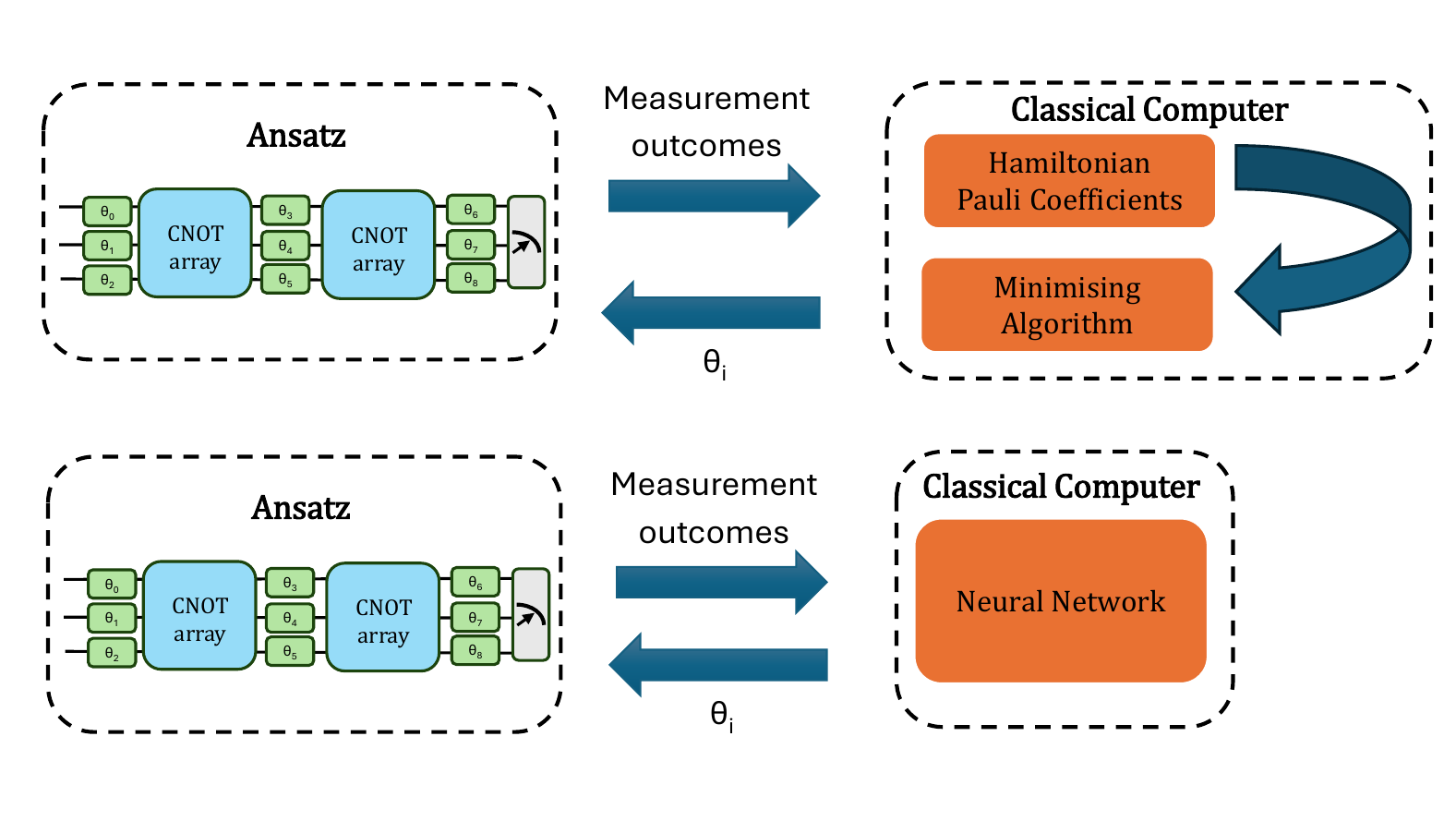}
        % Adjust coordinates as needed.
        \put(0,50){\textbf{a)}}
        \put(0,23){\textbf{b)}}
    \end{overpic}
    \caption{Illustration of machine learning optimisation. The quantum device is shown with the H$_2$ hardware efficient ansatz for three qubits. Initially in a), VQE is performed traditionally where the quantum component presents measurements to a classical computer which evaluates the loss function and runs a classical algorithm to predict the next set of parameters. Typically this loss function is also parameterised and the set of all intermediate measurement outcomes, final circuit parameters, and loss function parameters for a single VQE run for can be used to train the neural network. This neural network, shown in b), can then find optimal parameters for the circuit for different parameters of the loss function.}
    \label{fig:}
\end{figure*}

In this work, we investigate if the intermediate steps of VQE, which would naturally be generated by optimisation, can be used to train a machine learning model to find optimal circuit parameters for related Hamiltonians in few iterations. Figure~\ref{fig:} shows the full scheme, where a) are VQE optimisations performed initially then used as training for the machine learning model. Figure~\ref{fig:}b) then shows how this model can be used for further VQE runs. We show that the intermediate data from previous runs include information on the noise of the device and can not only allow machine learning to compensate for that specific device, but training across multiple noisy devices can allow machine learning to compensate for devices with arbitrary noise, thus recovering the noise resilience of VQE. Furthermore, we show a data augmentation scheme to reuse intermediate data is able to exponentially increase the training set for each new set of optimal parameters found.

To demonstrate the working of our technique, we investigate the potential energy surface of various Hydorgen and Helium based molecules with VQE and a feedforward deep neural network as the machine learning tool. We demonstrate resilience to not only simulated noise but noise on real IBM superconducting devices. We show that this neural network can be used as a replacement for classical optimisation and be capable of accelerating global optimisation while compensating for quantum processors with arbitrary coherent noise.

\section{Background}

For VQE, general purpose optimisers are typically used, of which COBYLA was found to be the fastest~\cite{Miki2022,Wibe2020}, however optimisers with better performance under noise have also been implemented~\cite{Wibe2020}. Similarly, parameters can be updated based on simulated imaginary time evolution~\cite{Stokes2020} or a Jacobi procedure with DIIS, similar to techniques used in classical computational chemistry~\cite{parrish2019}. In addition, specific optimisers for ADAPT-VQE have been designed, which are useful when the Ansatz needs to be constructed~\cite{armaos2021}. Likewise, there are optimisation strategies that use the structure of VQE parameter space to reduce overheads. Notably, the family of sequential optimisation strategies such as Rotosolve~\cite{Nakanishi2020,Ostaszewski2021}, Free Axis~\cite{Watanabe2021}, or Free Quaternion Selection~\cite{Wada2022}, which take precise measurements to fit the exact shape of the cost landscape and solve for exact minima along particular dimensions and has been found to scale better than stochastic optimisers and do not require hyperparameter tuning~\cite{Ostaszewski2021}. Modifications of the cost function have also been been found to speed up optimisation~\cite{PYin2020,Barkoutsos2020}. 

In addition to these, machine learning has found success in assisting optimisation for VQE. Simultaneous training of a VQE with a neural network was shown to speed up optimisation within one VQE run~\cite{Zhang2022} as well as mitigate barren plateaus~\cite{Lucas2022}. Reinforcement learning exhibited success on combinatorial problems, however this required exploratory runs on the system of interest to be optimised~\cite{Ostaszewski2024,Patel2024}.
Neural networks encoding Ansatz details have been shown to predict warm starts for larger Ansatze based on smaller Ansatze and this showed resilience to gaussian noise on circuit parameters~\cite{sauvage2021}. Furthermore, neural networks have been used exclusively as a noise mitigation step~\cite{Tran2024,Bennewitz2022,bhattacharjee2024}. Finally, neural networks have been used to directly learn the relationship between potential energy surface and molecular structure parameters using classical~\cite{Tao2022,Ghosh2023} and quantum neural networks~\cite{nishida2022} including compression with a quantum autoencoder~\cite{mesman2024}, however, effects of noise were not investigated. While these machine learning approaches have all found success, they all require training on separate test circuits or the VQE problem of interest and do not prove the ability to retain VQE noise resilience while reducing iterations required.

We demonstrate the ability to use intermediate steps of VQE calculations on molecular Hamiltonians with atoms at fixed distances to train a feedforward neural network. This neural network can then predict ground state energies of the molecule at arbitrary distances. We will now outline implementation details of VQE and the neural network.

\section{VQE Methodology}

Figure~\ref{fig:}a) shows an illustration of the VQE performed. We run VQE for the H$_2$ molecule reduced to one and two qubit Hamiltonians, H$_3$ linear chain molecule with three qubits, and HeH$^+$ with four qubits. These systems are typical for benchmarking of VQE methods and been extensively implemented in the literature~\cite{Akib2024, Peruzzo2014,Tilly2022}. We generate a Hamiltonian for the molecules at a fixed distance using PySCF~\cite{Sun2017} with a STO-3G basis~\cite{Pople2003}. For H$_2$, we map this to a one qubit basis using Jordan-Wigner mapping with qubit tapering~\cite{bravyi2017}, and a two qubit basis using the parity mapping~\cite{Love2012} with Z$_2$ symmetry reduction. For H$_3$ with three qubits, we use Jordan-Wigner mapping with qubit tapering and HeH$^+$ used Jordan-Wigner mapping with no tapering.

\begin{figure*}[htbp]
    \centering
    % Row 1
    \begin{minipage}[t]{0.45\textwidth}
        \begin{picture}(0,0)
            \put(-100,0){\makebox[0pt][l]{\textbf{a)}}}
        \end{picture}
        \includegraphics[width=\textwidth,valign=t]{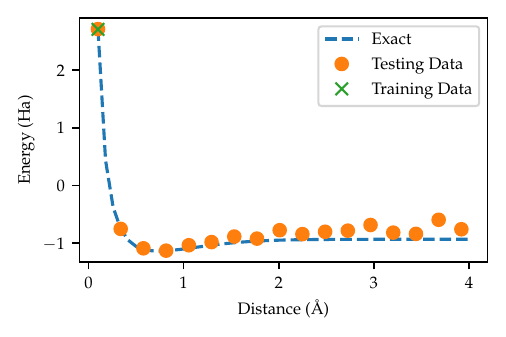}
    \end{minipage}
    \begin{minipage}[t]{0.45\textwidth}
        \begin{picture}(0,0)
            \put(-100,0){\makebox[0pt][l]{\textbf{b)}}}
        \end{picture}
        \includegraphics[width=\textwidth,valign=t]{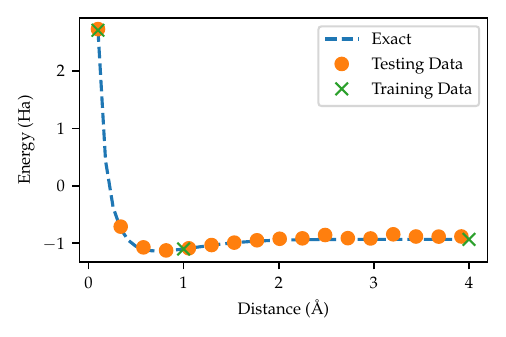}
    \end{minipage}
    
    \vspace{0.5cm}
    % Row 2
    \begin{minipage}[t]{0.45\textwidth}
        \begin{picture}(0,0)
            \put(-100,0){\makebox[0pt][l]{\textbf{c)}}}
        \end{picture}
        \includegraphics[width=\textwidth,valign=t]{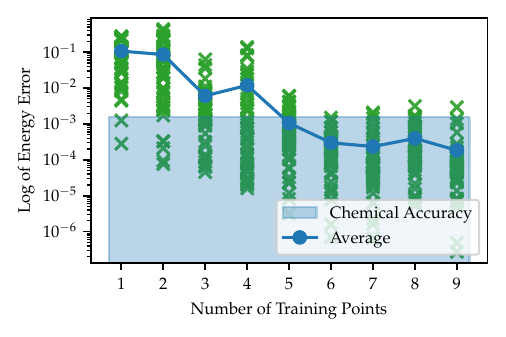}
    \end{minipage}
    \begin{minipage}[t]{0.45\textwidth}
        \begin{picture}(0,0)
            \put(-100,0){\makebox[0pt][l]{\textbf{d)}}}
        \end{picture}
        \includegraphics[width=\textwidth,valign=t]{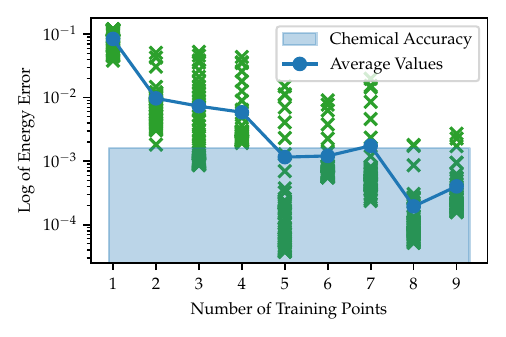} 
    \end{minipage}
    
    \caption{Training data required for neural network. VQE on one qubit H$_2$ was performed with COBYLA from random initial angles without noise for between one and nine equally spaced distances between the Hydrogen atoms. Similarly, VQE was performed on the four qubit HeH$^+$ system on the same distances. All intermediate steps were reused for each optimal parameter. The neural network was then tested on the entire curve by sampling fifty one equally spaced distances and beginning at random initial angles and iterating five times with the neural network and taking the minimum energy. a) shows the neural network predictions for one training point on seventeen of the fifty-one total testing points b) shows for three points. c) shows the error on a log scale across all fifty-one testing points as the training quantity increases. Average error is given as a blue dot and chemical accuracy is the shaded blue area. d) shows this same graph for the four qubit HeH$^+$ system.}
    \label{fig:train}
\end{figure*}
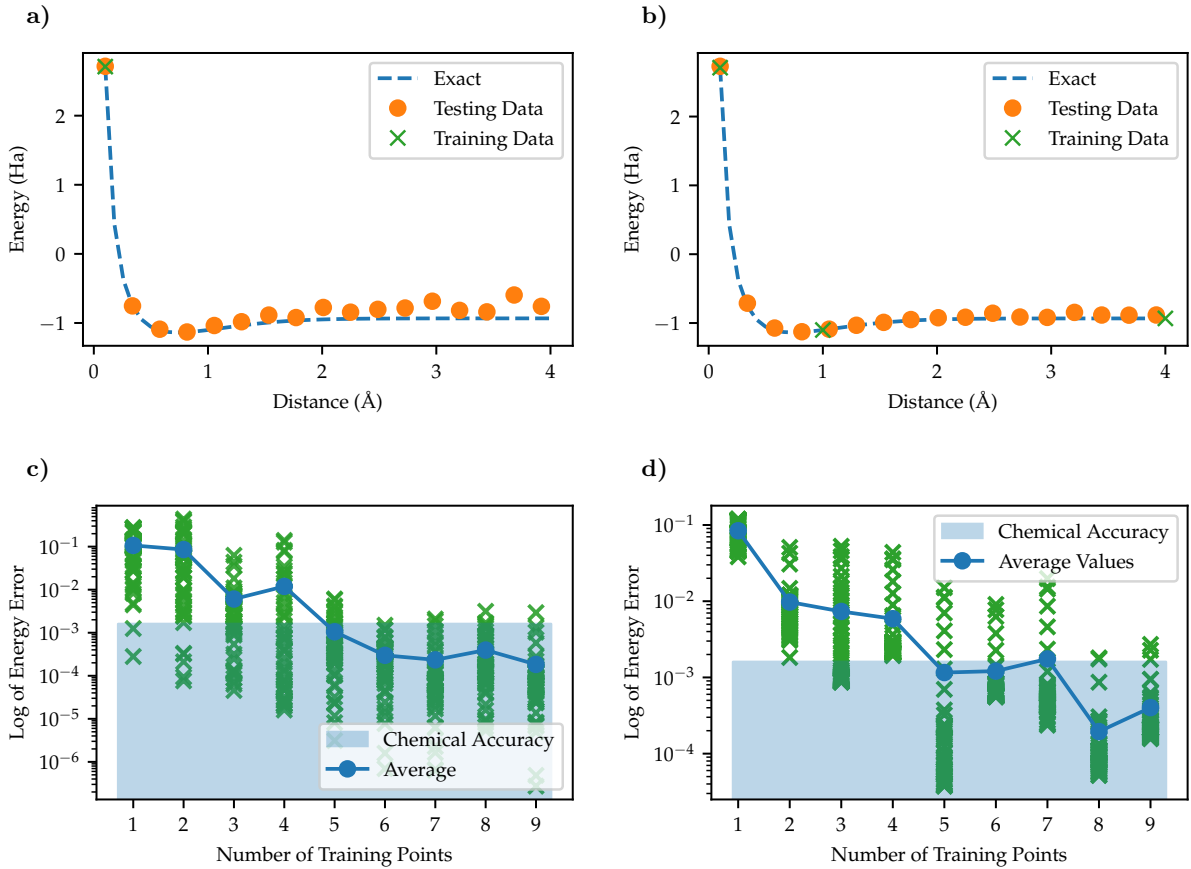

For the parametrised quantum circuit, we start with the Hartree-Fock state. For one qubit, we use a hardware efficient Ansatz. For two qubits we use a manually simplified UCCSD Ansatz with one parameter. For three qubits, we use a hardware efficient Ansatz, which has been shown to find the ground state~\cite{Akib2024}. An illustration of the hardware efficient circuit with two layers and three qubits is shown in Figure~\ref{fig:} as the Ansatz. For four qubits, we use a simplified UCCSD Ansatz by trotterising the doubles excitation before singles then manually simplifying based on the Hartree-Fock state. The exact circuits used are detailed in Appendix~\ref{app:i}. 

For this work, we choose COBYLA since it is likely to have fewer iterations than similar general use optimisers and more concentrated data than sequential optimisers such as Rotosolve~\cite{Nakanishi2020,Ostaszewski2021}. For sequential optimisers, intermediate steps measure a wide range of parameters, which would enrich the training set for the same number of iterations. We also choose to only use iterations up to a fixed cutoff rather than until convergence as the number of iterations to convergence varied significantly. Furthermore, the final step was included explicitly in the training set to encourage convergence. The machine learning models demonstrated should therefore be only improved by selecting alternative optimisers.

\section{Neural Network Training}

\begin{figure*}[htbp]
    \centering
    % Row 1
    \begin{minipage}[t]{0.40\textwidth}
        \begin{picture}(0,0)
            \put(-100,0){\makebox[0pt][l]{\textbf{a)}}}
        \end{picture}
        
    \hspace{-0.3cm}
        \includegraphics[width=\textwidth,valign=t]{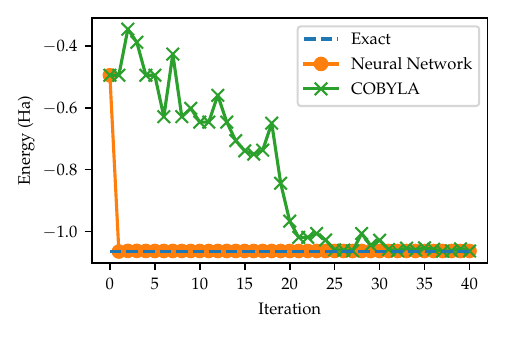}
    \end{minipage}
    \hspace{0.5cm}
    \begin{minipage}[t]{0.45\textwidth}
        \begin{picture}(0,0)
            \put(-100,0){\makebox[0pt][l]{\textbf{b)}}}
        \end{picture}
        \includegraphics[width=\textwidth,valign=t]{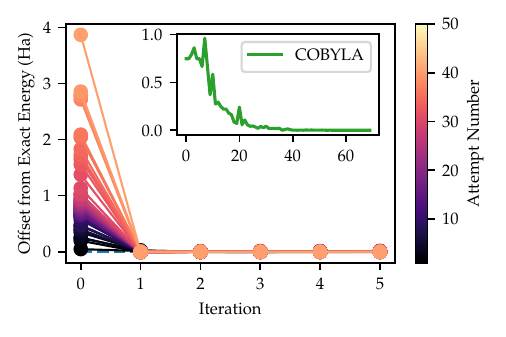}
    \end{minipage}
    
    \vspace{0.5cm}
    % Row 2
    \begin{minipage}[t]{0.45\textwidth}
        \begin{picture}(0,0)
            \put(-100,0){\makebox[0pt][l]{\textbf{c)}}}
        \end{picture}
        \includegraphics[width=\textwidth,valign=t]{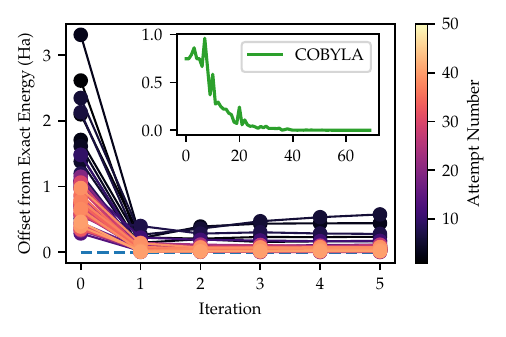}
    \end{minipage}
    \begin{minipage}[t]{0.45\textwidth}
        \begin{picture}(0,0)
            \put(-100,0){\makebox[0pt][l]{\textbf{d)}}}
        \end{picture}
        \includegraphics[width=\textwidth,valign=t]{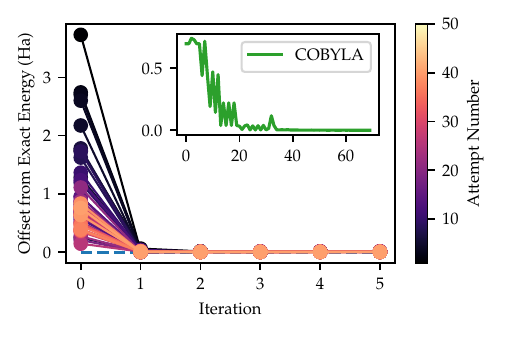}
    \end{minipage}

       \caption{Effect of noise on the neural network. The system is the H$_2$ Hamiltonian for four qubits with hardware efficient Ansatz with 8 parameters. a) shows the performance of the neural network against COBYLA, a standard optimisation algorithm with noise-free training on a noise-free simulated device with convergence of $10^{-2}$ Hartree. b) shows this neural network applied to 30 random Hamiltonians. The attempts are sorted in order of furthest initial angle from the ideal. c) shows the neural network from a) running on 30 devices with fixed over/under-rotation errors. d) shows the neural network trained on devices with a random fixed over/under-rotation error between $-0.1 \pi$ to $0.1 \pi$}
    \label{fig:noise}
\end{figure*}

Figure~\ref{fig:}b) shows the neural network replacing the minimising algorithm. It takes the Hamiltonian as a Pauli vector with the angles and measurement outcomes of the quantum circuit as input and outputs an update to the angles. 

For initial testing, we run VQE with COBYLA for a varying number of points along the potential energy surface and test how well the neural network can predict the rest of the surface. For the following data, COBYLA was initialised from random angles and the first $10$ steps for H$_2$ with one qubit and $30$ steps for HeH$^+$ with four qubits were used as training to facilitate comparison. All training data was simulated for this work. The final angles with expectation values were also included in the data set.

The machine learning algorithm consists of a feed forward neural network with ReLU activation function with four layers and a number of neurons starting at the input size descending to the output size, typically by halving the number of neurons in each layer. The input vector for the neural network are the Hamiltonian coefficients for each Pauli string, which represent the loss function, the angles for the Ansatz, and corresponding expectation values for each Pauli string. The output vector are the optimal minimal angles. For training, the optimal angles were found analytically for one qubit as described in Appendix~\ref{app:ii} or convergence with COBYLA for two, three, and four qubit systems.

Furthermore, data augmentation is performed by reusing intermediate measurements in the training set. The input vector consists of: the measured angle, the quantum circuit measurement outcomes, and the Hamiltonian Pauli vector. The output is the difference with the final angle, so for each final angle found, the Hamiltonian Pauli vector and angle difference can be easily generated and the same angle and measurements can be reused. This allows each VQE run to more than double the training dataset, resulting in exponential growth of training data.

The training data was generated from each intermediate COBYLA step. The expectation values, the measured angle, and the Hamiltonian Pauli vector were concatenated as inputs, and the difference between the measured angle and the optimal angle was given as output. For testing, 1\% of this data was separated. 

First, the neural network was initialised using He initialisation~\cite{He2015}. Training occurred for 1000 epochs using mini-batches with batch size of $2^3$. All neural networks reached a mean squared error on training data of less than $10^{-4}$ and the testing data from COBYLA of less than $10^{-3}$. The neural networks were then tested by predicting Hamiltonians on equally spaced distances between Hydrogen atoms.

Figure~\ref{fig:train} shows the results for varying number of training points on the one qubit H$_2$ system. With one training point at very close distances, the neural network is able to more accurately predict Hamiltonians neighbouring the testing point and worsens for Hamiltonians corresponding to longer bond lengths. From Figure~\ref{fig:train}b), we can see that angles for Hamiltonians neighbouring the training data are still accurately predicted but the energy for bond lengths around $2$ \AA~are still inaccurate as in Figure~\ref{fig:train}a). While previous work encodes a one dimensional distance parameter as machine learning input~\cite{Ghosh2023} to predict one direction of the potential energy surface, a neural network is able to learn the full loss function, which will allow the same neural network to consider arbitrary parametrisations of the Hamiltonians, for instance varying angles rather than distances, which are more useful for constrained optimisations or adsorption studies.

Figure~\ref{fig:train}c) shows all 51 testing points for one qubit H$_2$ as the number of evenly spaced training data points increases on a log plot of difference between predicted energy and the exact energy. Similarly, d) shows the same graph for four qubit HeH$^+$. We also show the full potential energy surface for H$_2$ for one training data point in a) and three in b). Training loss reached order $10^{-4}$ for all cases. Hyperparameters for all training were 1000 epochs with $2^3$ batch size for one qubit H$_2$ and $2^7$ batch size for HeH$^+$. Tuning of hyperparameters for each training set may result in better performance.

From Figure~\ref{fig:train}, we can see that increasing the number of points increases the average accuracy of the full curve. At $5$ data points, the average of points lie within chemical accuracy for both H$_2$ and HeH$^+$. Furthermore, in actual calculations, chemical accuracy from direct neural network measurements may not be necessary, for instance, in previous work~\cite{sauvage2021}, these machine learning predictions were used for warm starts with subsequent gradient descent to more accurate values. However, there also exist techniques such as the moments method~\cite{Jones2022}, which are able to use additional measurements to find accurate ground states from neighbouring states. In that sense, a full application may only require a few VQE runs to initially train the neural network. Once it is trained, the neural network can predict angles to measure for related Hamiltonians and only single measurements with corrections such as the moments method are required to find the ground state within chemical accuracy.

\section{Comparison under Simulated Noise}

We now compare the performance of the neural network predictions with COBYLA in noise free and noisy conditions. For this we use the four qubit HeH$_+$ system. Training occurred with 5 Hamiltonians equally spaced along the potential energy surface with COBYLA run until convergence from a random initial starting point. For noise free conditions, batch sizes of $2^{4}$ were used. 

A sample individual testing run can be seen in Figure~\ref{fig:noise}a). COBYLA searches for the minimum by exploring the space before converging, whereas the neural network can find the optimal angles in one shot but may oscillate around the angle. COBYLA took 50 iterations to converge, so Figure~\ref{fig:noise}a) was cutoff at 40 for comparison. The neural network appears to converge, however there is no mathematical guarantee. The training data generated by COBYLA has a higher concentration of data around the optimal angle which may assist in the appearance of convergence. For this work, we restricted the number of COBYLA datapoints to the first 30, however the final converged measurement was added manually to encourage convergence.

For Figure~\ref{fig:noise}b), we run the neural network and COBYLA algorithms from a) for random initial starting points on 5 random Hamiltonians. As with Figure~\ref{fig:train}d), the neural network is able to start from arbitrary initial positions and find the optimal angles. In most cases, the optimal angle is found in one iteration before oscillating, whereas a few runs required two iterations. 

We now test the neural network under simulated noise models. We consider noise to be coherent noise that can be compensated for with adjustment to the Ansatz angles. Most obviously, over/under rotation of the angles in hardware will contribute to this noise, however it is also possible that some processes such as thermal relaxation and error channels such as amplitude damping cause excited or $\ket{1}$ states to rotate to $\ket{0}$, and adjusting the angles for the VQE Ansatz may be able to compensate, however this depends heavily on the state and the Ansatz. For this work, we model fixed over/under rotation angle for each Ansatz angle. Since this is effectively a static shift on the Ansatz angles, it makes no difference to either COBYLA or the neural network and performance is identical to Figure~\ref{fig:noise}. 

However, we want to consider multiple devices with different fixed over/under rotation errors. This simulates drift or an experiment that occurs on one device across multiple days or with interruption or calibration. In this case, we simulate 30 random errors on each angle between $-0.1 \pi$ to $0.1 \pi$ radians. Figure~\ref{fig:noise}c) shows the neural network trained on noise-free data from Figure~\ref{fig:noise}a) run on these devices. Without training from the device, the neural network can only predict angles for a noise free device, which can be seen by the oscillation around energies much higher than the exact. COBYLA can compensate for each rotation error separately, but will be need a fresh run each time and has no mechanism to learn between devices. Furthermore, it cannot converge within 5 iterations.

\begin{figure*}[htbp]
    \centering
    % Row 1
    \begin{minipage}[t]{0.45\textwidth}
        \begin{picture}(0,0)
            \put(-100,0){\makebox[0pt][l]{\textbf{a)}}}
        \end{picture}
        \includegraphics[width=\textwidth,valign=t]{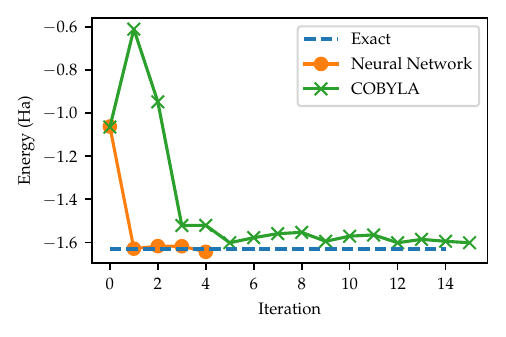}
    \end{minipage}
    \begin{minipage}[t]{0.45\textwidth}
        \begin{picture}(0,0)
            \put(-100,0){\makebox[0pt][l]{\textbf{b)}}}
        \end{picture}
        \includegraphics[width=\textwidth,valign=t]{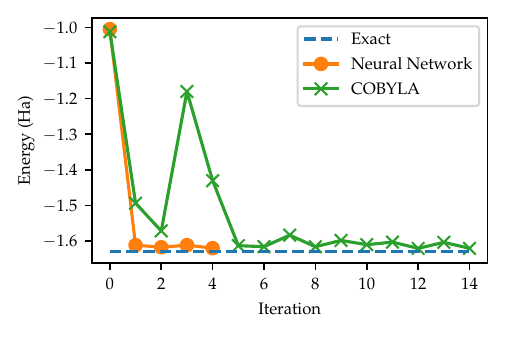}
    \end{minipage}
    
    \vspace{0.5cm}
    % Row 2
    \begin{minipage}[t]{0.45\textwidth}
        \begin{picture}(0,0)
            \put(-100,0){\makebox[0pt][l]{\textbf{c)}}}
        \end{picture}
        \includegraphics[width=\textwidth,valign=t]{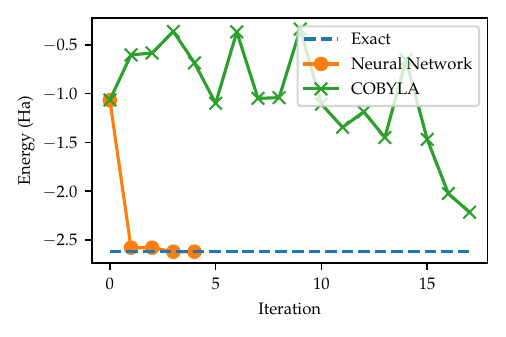}
    \end{minipage}
    \begin{minipage}[t]{0.45\textwidth}
        \begin{picture}(0,0)
            \put(-100,0){\makebox[0pt][l]{\textbf{d)}}}
        \end{picture}
        \includegraphics[width=\textwidth,valign=t]{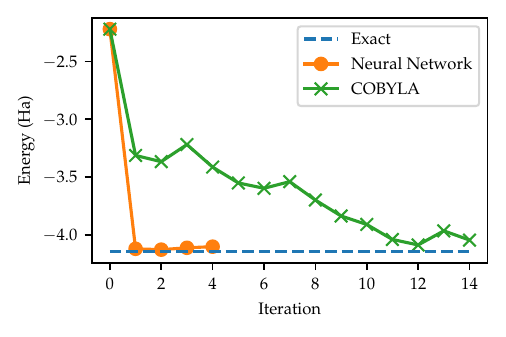}
    \end{minipage}
        \caption{VQE performance on IBM superconducting device $ibm\_brisbane$ for one to four qubit systems. Exact refers to direct diagonalisation of the Hamiltonian. Neural network refers to the network fully trained on all error combinations and applied on real machine data for 5 iterations. COBYLA refers to the classical algorithm used on real machine data for 15 iterations. a) shows the one qubit H$_2$ molecule; b) shows the H$_2$ molecule for two qubits with modified UCC Ansatz; c) shows the H$_3$ linear chain molecule with hardware efficient Ansatz for three qubits. d) shows HeH$^+$ with UCCSD Ansatz for four qubits. Noise mitigation was performed on c) and d) by dividing by reference state measurements of $0.96$ and $0.98$ respectively to remove depolarisation noise based on a test measurement with all parameters set to zero.}
    \label{fig:exp}
\end{figure*}
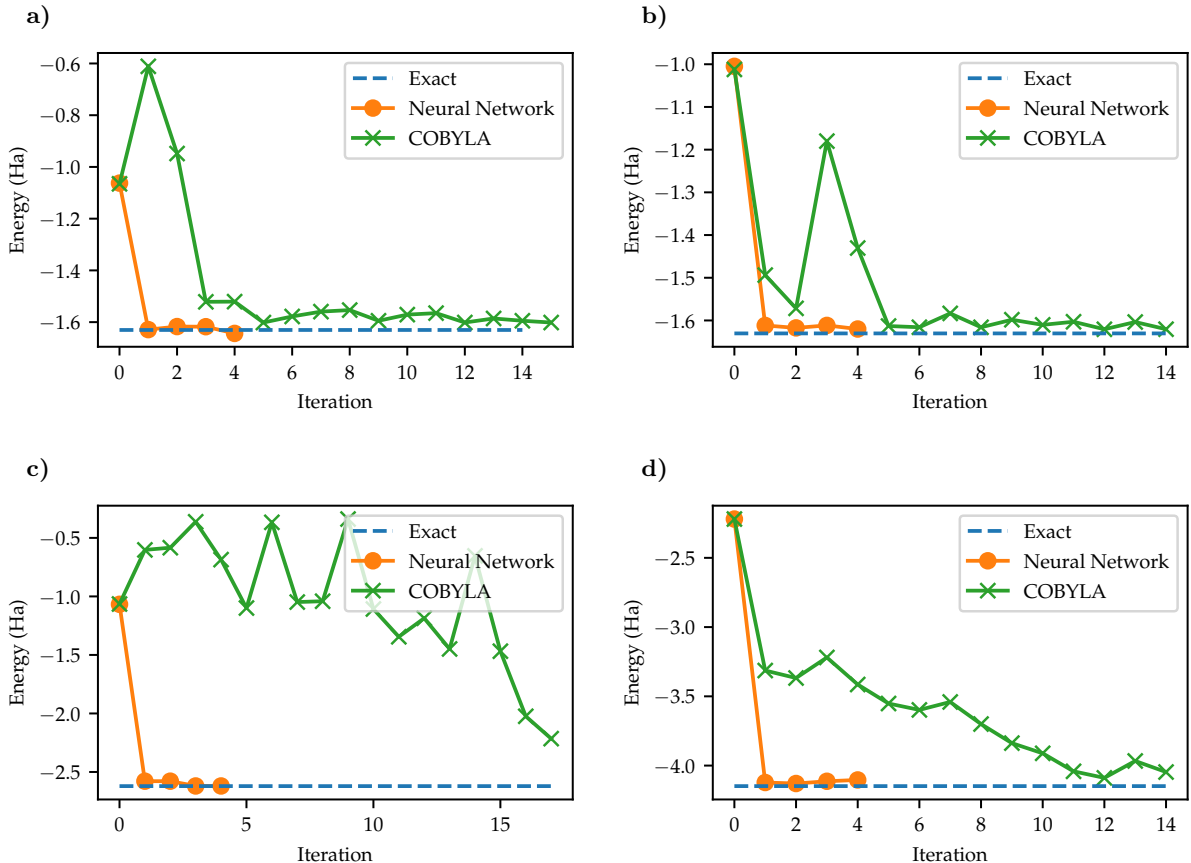

Figure~\ref{fig:noise}d) shows the neural network trained with the same data as a) but modified with noise. The VQE runs are modified for each combination of over and under rotation errors for each angle and the total set is used for training data. Due to the size of the training data, batch size of $2^{11}$ were used for minibatch training. The neural network is now able to compensate for any noise in the neighbourhood of the devices it was trained on.
\section{Performance on IBM Quantum Devices}

In this section, the neural networks trained on all combinations of over/under-rotation are run on a real device. No retraining on devices was performed and all neural networks use simulated data for training.

We run the one qubit H$_2$ system, a two qubit version of the H$_2$ system with Parity mapping and symmetry reduction, the three qubit H$_3$ system, and four qubit HeH$^+$. All results are obtained from $ibm\_brisbane$ shown in Figure~\ref{fig:exp}a) with $10,000$ shots to minimise shot noise. Due to COBYLA taking an unknown number of steps, the calculations were truncated. We disabled the inbuilt optimisation and noise mitigation methods by IBM.

The inbuilt IBM noise mitigation was avoided since it may correct the coherent errors. Our method will compensate for coherent noise but not incoherent noise, so we implemented a basic correction for only incoherent noise. The simplest model of incoherent noise is the totally depolarising error channel, so we perform noise mitigation by first removing all the rotation gates in the Ansatz. This will leave a circuit composed of only Clifford gates that prepares a state with ZZZ expectation value of either -1 or 1. Therefore, assuming total depolarising channels across each two qubit gate, the measured expectation value will be some depolarising parameter $p$ up to a sign. Once this is evaluated, the remaining Pauli string expectation values can be divided by $p$. This can be seen as a variation of reference state error mitigation that should compensate for incoherent depolarisation based errors but not coherent errors.

The final results are given in Figure~\ref{fig:exp}. The experimental values of $p$ were $0.96$ for three qubits, $0.98$ for four qubits, and above $1$ for one and two qubits, which indicates the shot noise was higher than the depolarising noise, consistent with the low depth Ansatze compared to three and four qubits. Therefore, one and two qubits were not corrected for incoherent noise. For one and two qubits, COBYLA was run for 15 steps and did not converge within $10^{-4}$ Hartree tolerance but reached within $10^{-1}$ Hartree. Similarly for three qubits, COBYLA was run for 18 steps but did not converge and did not reach within $10^{-1}$ Hartree and for four qubits, COBYLA was run for 25 steps and reached within $10^{-2}$ Hartree.

We find in all cases that the neural network is able to successfully find the neighbourhood of an optimal angle in one shot and then oscillate around the optimal angle. Furthermore, on a real device, there is shot noise causing further noisy fluctuations. The angle oscillations will cause the value to be higher than the exact energy due to the variational principle, however shot noise can cause the energy to be lower. This is seen in the one qubit case in Figure~\ref{fig:exp}(a). There will be no shot noise for the two qubit case in Figure~\ref{fig:exp}b), so the oscillations are due to the neural network alone and all energies are at or above the exact value. For three qubits in c), the lowest value is $-2.61955$ Ha compared to the exact value of $-2.62078$ Ha. While this appears close, shot noise is able to violate the variational principle and reduce the energy below the exact value, so we suspect the noise reduction by assuming a total depolarisation channel is unable to fully account for the noise and the shot noise fluctuation appears to compensate for the remaining increase in energy. Similarly, d) we are able to reach $-4.12951$ Ha compared to exact $-4.14827$ Ha which is lower, likely due to shot noise both in the measurement and the reference state measurement used for correction.

All the neural networks were able to compensate for any over/under-rotations on the real device. This is most likely to occur in the three and four qubit cases, due to the deeper Ansatz and the lack of optimisations for noise. The success of these results in Figure~\ref{fig:exp}c) and d) demonstrate that a neural network trained on virtual data of combinations of over/under-rotation can successfully compensate for noise on a real device.

While we have demonstrated the technique on a one dimensional parametrisation of the geometry, this can be easily applied to other parametrisations due to the Hamiltonian being encoded as input. This allows the same neural network to work independent of the geometric parametrisation. For instance, the H$_3$ linear chain was parametrised by one distance parameter with equal spacing between atoms, however Hamiltonians from other geometries like a torsional study with rotation axis, could be added to the training data and the same neural network architecture can be used. Similarly, for solid state, bandstructures or density of states can be found by sampling from the reciprocal space~\cite{Shaobo2024}, which requires multiple VQE runs and would be suitable to use this technique with. Finally, this technique is suitable for any variational algorithm where different loss functions need to be studied for a fixed Ansatz. Modification will be required for example, for QAOA which changes the Ansatz depending on the loss function. Similarly, the method may be compatible with collective optimisation~\cite{PYin2020}, which can also be used for potential energy surface determination along a path. Our technique can be modified to act as an optimiser if multiple paths in Hamiltonian space are required; or give a warm start to a collective optimisation if one initial VQE is performed.

\section{Conclusion}

We demonstrate that machine learning is able to use intermediate steps of previous VQE runs to extract relationships between the Hamiltonian loss function and the optimal angles as well as coherent noise on a device. We show that a neural network trained on noisy devices is able to predict optimal parameters for other noisy devices. Furthermore, this works on real devices and is able to find the optimal angles in one shot in noisy environments. At scale, this shows that intermediate VQE steps, including expectation values, should be stored for machine learning. Furthermore, this is valuable for devices that have interrupted access or drift where it can reduce the required VQE steps or even find optimal angles in one shot for sufficient available training data.

\section{Acknowledgements}
The research was supported by the University of Melbourne through the establishment of the IBM Quantum Network Hub at the University. The authors would like to thank Michael A. Jones for discussions on ansatz design.

\bibliography{abstract}

\appendix
\section{Ansatze Circuits}\label{app:i}
In this section we will provide the circuits run on real devices for each system. For one qubit, we used a hardware efficient Ansatze; for two qubits we used UCCSD that was manually modified~\cite{Akib2024}; for three qubits we used a hardware efficient Ansatze; and for four qubits w used UCCSD that was manually modified for this work.

For HeH$^+$, we consider STO-3G basis set which models one $s$ orbital per atom, resulting in a four orbital basis of two spin up and two spin down orbitals. It should be noted that these are not the atomic orbitals localised to each atom but the molecular orbitals solved by Hartree-Fock. The Hartree-Fock initial guess for two electrons after Jordan-Wigner mapping is $\ket{0101}$. We use qiskit notation where the first two are spin $\alpha$ and the last two are spin $\beta$.

For UCCSD, we consider only particle preserving and spin preserving excitations. For $a_i$ as fermionic annihilation operator and $a_i^\dagger$ as creation operator for state $i$, this leaves us with two single excitation operators:

\begin{align}
    T_{s1} = a_0^\dagger a_1,\\
    T_{s2} = a_2^\dagger a_3.
\end{align}

Double excitation is similarly given by:

\begin{equation}
    T_d = a_0^\dagger a_2^\dagger a_1 a_3.
\end{equation}

These are mapped to qubit space using the Jordan-Wigner transformation:

\begin{align}
a_j &= \left( \prod_{k=1}^{j-1} Z_k \right) \frac{1}{2}\Bigl(X_j - i\, Y_j\Bigr),\\[1ex]
a_j^\dagger &= \left( \prod_{k=1}^{j-1} Z_k \right) \frac{1}{2}\Bigl(X_j + i\, Y_j\Bigr).
\end{align}

For instance, the first singles operator becomes $T_{s1} = iXYII$.

Now the full UCCSD Ansatz acting on the Hartree-Fock state is given by:

\begin{equation}
    e^{\frac{1}{2}(t_0T_{s1} + t_1T_{s2} + t_3T_d - h.c.)}\ket{0101},
\end{equation}
where $t$ are parameters to optimise and h.c. is the Hermitian conjugate required so that the exponential forms a unitary matrix.

We then Trotterize this with one step to give:
\begin{equation}
    e^{\frac{t_0}{2}(T_{s1}-h.c.)}e^{\frac{t_1}{2}(T_{s2}-h.c.)}
    e^{\frac{t_2}{2}(T_{d} - h.c.)}\ket{0101}.
\end{equation}

\begin{figure*}
    \centering\scalebox{1.0}{
\Qcircuit @C=1.0em @R=0.2em @!R { \\
	 	\nghost{{q}_{0} :  } & \lstick{{q}_{0} :  } & \gate{\mathrm{X}} & \gate{\mathrm{H}} & \qw & \ctrl{1} & \gate{\mathrm{S^\dagger}} & \gate{\mathrm{R_Y}\,(\mathrm{\frac{{\ensuremath{\theta}}[2]}{4}})} & \ctrl{1} & \gate{\mathrm{H}} & \qw & \qw\\
	 	\nghost{{q}_{1} :  } & \lstick{{q}_{1} :  } & \gate{\mathrm{R_Y}\,(\mathrm{\frac{{\ensuremath{\theta}}[0]}{2}})} & \ctrl{1} & \gate{\mathrm{S^\dagger}} & \targ & \gate{\mathrm{S}} & \gate{\mathrm{R_Y}\,(\mathrm{\frac{{\ensuremath{\theta}}[2]}{4}})} & \targ & \qw & \qw & \qw\\
	 	\nghost{{q}_{2} :  } & \lstick{{q}_{2} :  } & \gate{\mathrm{X}} & \targ & \gate{\mathrm{S^\dagger}} & \targ & \gate{\mathrm{S}} & \gate{\mathrm{R_Y}\,(\mathrm{\frac{{\ensuremath{\theta}}[1]}{4}})} & \targ & \qw & \qw & \qw\\
	 	\nghost{{q}_{3} :  } & \lstick{{q}_{3} :  } & \gate{\mathrm{X}} & \gate{\mathrm{H}} & \qw & \ctrl{-1} & \gate{\mathrm{S^\dagger}} & \gate{\mathrm{R_Y}\,(\mathrm{\frac{{\ensuremath{\theta}}[1]}{4}})} & \ctrl{-1} & \gate{\mathrm{H}} & \qw & \qw\\
\\ }}
\caption{Manually simplified four qubit UCCSD Ansatz for HeH$^+$. Initial Hartree-Fock state is mapped to $\ket{0101}$. Doubles are Trotterised before singles and replaced with the circuit that maps the Hartree-Fock state to a superposition with the doubly excited state.}\label{fig:hehqc}
\end{figure*}
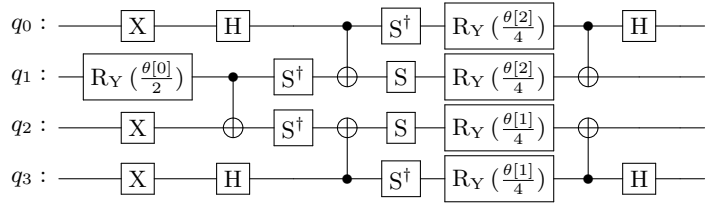

We do not have to design a circuit that implements the full unitary for the doubles term, only its operation on the Hartree-Fock state. Specifically, the doubles term applied to the state is:
\begin{equation}
    e^{\frac{t_2}{2}(T_d - h.c.)} \ket{0101} = \cos(t_2)\ket{0101}+\sin(t_2)\ket{1010},
\end{equation} so we only have to design a circuit that prepares this state. Doing this and adding the two single excitation circuits for qubit 0 and 1; and qubit 2 and 3 with basic simplification of gates, results in the circuit in Figure~\ref{fig:hehqc}.

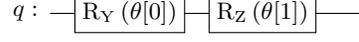
\begin{figure*}
    \centering\scalebox{1.0}{
\Qcircuit @C=1.0em @R=0.2em @!R { \\
	 	\nghost{{q} :  } & \lstick{{q} :  } & \gate{\mathrm{R_Y}\,(\mathrm{\theta}[0])} & \gate{\mathrm{R_Z}\,(\mathrm{\theta}[1])} & \qw & \qw\\
\\ }}
\caption{Hardware efficient one qubit circuit for H$_2$ one qubit Hamiltonian.}\label{fig:oneq}
\end{figure*}

\begin{figure*}
    \centering
\scalebox{1.0}{
\Qcircuit @C=1.0em @R=0.2em @!R { \\
	 	\nghost{{q}_{0} :  } & \lstick{{q}_{0} :  } & \gate{\mathrm{R_Y}\,(\mathrm{\theta})} & \ctrl{1} & \qw & \qw\\
	 	\nghost{{q}_{1} :  } & \lstick{{q}_{1} :  } & \gate{\mathrm{X}} & \targ & \qw & \qw\\
\\ }}
    \caption{Manually reduced Ansatz for Z$_2$ symmetry reduced UCC H$_2$ two qubit Hamiltonian. Details in Karim et al.~\cite{Akib2024}. The Hartree Fock initial guess is mapped to $\ket{10}$. The excitations were found to map to the space spanned by $\ket{01}$ and $\ket{10}$ with real coefficients.}
    \label{fig:app1}
\end{figure*}
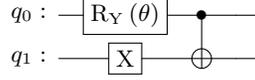

\begin{figure*}
    \centering\scalebox{0.7}{
\Qcircuit @C=1.0em @R=0.2em @!R { \\
	 	\nghost{{q}_{0} :  } & \lstick{{q}_{0} :  } & \gate{\mathrm{R_Y}\,(\mathrm{{\ensuremath{\theta}}[0]})} & \gate{\mathrm{R_Z}\,(\mathrm{{\ensuremath{\theta}}[3]})} & \qw & \qw & \ctrl{1} & \gate{\mathrm{R_Y}\,(\mathrm{{\ensuremath{\theta}}[6]})} & \gate{\mathrm{R_Z}\,(\mathrm{{\ensuremath{\theta}}[9]})} & \qw & \ctrl{1} & \gate{\mathrm{R_Y}\,(\mathrm{{\ensuremath{\theta}}[12]})} & \gate{\mathrm{R_Z}\,(\mathrm{{\ensuremath{\theta}}[15]})} & \qw & \qw\\
	 	\nghost{{q}_{1} :  } & \lstick{{q}_{1} :  } & \gate{\mathrm{X}} & \gate{\mathrm{R_Y}\,(\mathrm{{\ensuremath{\theta}}[1]})} & \gate{\mathrm{R_Z}\,(\mathrm{{\ensuremath{\theta}}[4]})} & \ctrl{1} & \targ & \gate{\mathrm{R_Y}\,(\mathrm{{\ensuremath{\theta}}[7]})} & \gate{\mathrm{R_Z}\,(\mathrm{{\ensuremath{\theta}}[10]})} & \ctrl{1} & \targ & \gate{\mathrm{R_Y}\,(\mathrm{{\ensuremath{\theta}}[13]})} & \gate{\mathrm{R_Z}\,(\mathrm{{\ensuremath{\theta}}[16]})} & \qw & \qw\\
	 	\nghost{{q}_{2} :  } & \lstick{{q}_{2} :  } & \gate{\mathrm{R_Y}\,(\mathrm{{\ensuremath{\theta}}[2]})} & \gate{\mathrm{R_Z}\,(\mathrm{{\ensuremath{\theta}}[5]})} & \qw & \targ & \gate{\mathrm{R_Y}\,(\mathrm{{\ensuremath{\theta}}[8]})} & \gate{\mathrm{R_Z}\,(\mathrm{{\ensuremath{\theta}}[11]})} & \qw & \targ & \gate{\mathrm{R_Y}\,(\mathrm{{\ensuremath{\theta}}[14]})} & \gate{\mathrm{R_Z}\,(\mathrm{{\ensuremath{\theta}}[17]})} & \qw & \qw & \qw\\
\\ }}
    \caption{Three qubit two layer Hardware efficient Ansatz for the Jordan-Wigner H$_3$ Hamiltonian. The initial Hartree-Fock guess is mapped to $\ket{010}$.}
    \label{fig:app2}
\end{figure*}
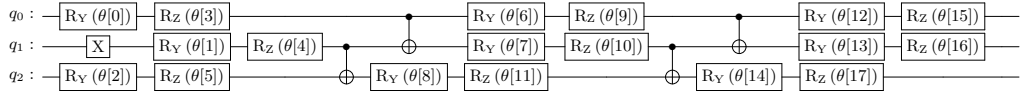

\section{Analytical One Qubit Solution}\label{app:ii}

We derive the analytical solution for the optimal ansatz angles for minimising the energy of a one qubit Hamiltonian using the Ansatz in Figure~\ref{fig:oneq}. 

Consider the one qubit Ansatz with angles $u$ for $R_y$ and $v$ for $R_z$. This prepares the state:
\begin{equation}
\left[\begin{matrix}e^{- 0.5 i v} \cos{\left(\frac{u}{2} \right)}\\e^{0.5 i v} \sin{\left(\frac{u}{2} \right)}\end{matrix}\right]
\end{equation}

The expectation value of $X$ is $\sin(u)\cos(v)$; $Y$ is $\sin(u)\sin(v)$; and $Z$ is $\cos(u)$.

Now consider a Hamiltonian written as a Pauli vector: 
\begin{equation}
    H = h_0I + h_1X + h_2Y + h_3Z,
\end{equation}

We omit $h_0$ as it only offsets the energy. We then substitute:
\begin{equation}
    a e^{ib} = h_1 + ih_2,
\end{equation}

so the Hamiltonian we use is now:
\begin{equation}
    H_{opt} = \frac{a}{2}(e^{-ib}(X+iY) + e^{ib}(X-iY)) + h_3Z
\end{equation}

The expectation value of the Ansatz state with this Hamiltonian is given by:

\begin{equation}
\braket{H_{opt}} = a\sin(u)\cos(b-v) + h_3\cos(u)
\end{equation}

There is a trivial solution of $-h_3$ for $u = \pi$ or $h_3$ for $u=0$ if $h_3 < 0$ with arbitrary $v$. We see there is a minimum energy of $-\sqrt{h_3^2 + a^2}$ for:

\begin{equation}
    u_{opt} = 2\tan^{-1}\left( \frac{h_3 + \sqrt{h_3^2 + a^2}}{a} \right), v_{opt}=b.
\end{equation}

This gives a final energy of:
\begin{equation}
    \braket{H} = h_0-\sqrt{h_1^2 + h_2^2 + h_3^2 }
\end{equation}.

For neural network training, the input vector is the Hamiltonian parameters $h_1,h_2,h_3$ with corresponding expectation values $\sin(u)\cos(v), \sin(u\sin(v), \cos(u)$, and then angles $u,v$. The output are the correction to the angles to find the final angles: $u_{opt} - u, v_{opt}-v$.

In the case of over/under rotation noise, the expectation values will be modified by some $u_{err}$ and $v_{err}$:

\begin{equation}
    \begin{pmatrix}
    h_1 \\
    h_2 \\
    h_3 \\
    \sin(u + u_{err}) \cos(v + v_{err}) \\
    \sin(u + u_{err}) \sin(v + v_{err}) \\
    \cos(u + u_{err})
    \end{pmatrix}
\end{equation}
and the corresponding output vector will be:

\begin{equation}
    \begin{pmatrix}
    u_{opt} - u - u_{err} \\
    v_{opt} - v - v_{err}
    \end{pmatrix}
\end{equation}.

Expanded in terms of Hamiltonian coefficients this is:
\begin{equation}
    \begin{pmatrix}
    2\tan^{-1}\left( \frac{h_3 + \sqrt{h_1^2 + h_2^2 + h_3^2}}{\sqrt{h_1^2+h_2^2}} \right) - u - u_{err} \\
    \tan^{-1}(\frac{h_2}{h_1}) - v - v_{err}
    \end{pmatrix}
\end{equation}.

This mapping is what the neural network finds.

\end{document}